\documentstyle[prl,aps,graphicx,floats]{revtex}
\begin{document}
\draft
\title{
Evidence for Linelike Vortex Liquid Phase in Tl$_2$Ba$_2$CaCu$_2$O$_8$ Probed by
the Josephson Plasma Resonance}
\author{V. K. Thorsm{\o}lle$^1$, R. D. Averitt$^1$, M. P. Maley$^1$, M. F. Hundley$^1$,
A. E. Koshelev$^2$, L. N. Bulaevskii$^1$, and~A.~J.~Taylor$^1$}
\address{$^1$Los Alamos National Laboratory, Los Alamos, NM 87545}
\address{$^2$Materials Science Division, Argonne National Laboratory, Argonne, IL 60469}
\date{\today}

\wideabs{ %
\maketitle
\begin{abstract}
We measured the Josephson plasma resonance (JPR) in optimally doped
Tl$_2$Ba$_2$CaCu$_2$O$_{8+\delta}$ thin films using terahertz time-domain
spectroscopy in transmission.  The temperature and magnetic field
dependence of the JPR frequency shows that the $c$-axis correlations of
pancake vortices remain intact at the transition from the vortex solid to
the liquid phase.  In this respect Tl$_2$Ba$_2$CaCu$_2$O$_{8+\delta}$ films, with
anisotropy parameter $\gamma\approx 150$, are similar to the less
anisotropic YBa$_2$Cu$_3$O$_{7-\delta}$ $(\gamma\approx 8)$ rather than to
the most anisotropic Bi$_2$Sr$_2$CaCu$_2$O$_{8+\delta}$ single crystals ($\gamma\geq
500$).
\end{abstract}
 \vspace*{-10pt}}  \narrowtext

The structure of the different vortex phases and the nature of the phase
transitions in the magnetic phase diagram of high-$T_c$ superconductors
has been the subject of intense study over the past several years
\cite{Blatteretal}.  It has been established rather conclusively, both
experimentally \cite{Zeldovetal,Schillingetal} and theoretically, that
there exists a first-order phase transition at which the vortex lattice
melts into a vortex liquid where at least in-plane long-range order is
lost.  This melting transition depends strongly on the anisotropy
parameter, $\gamma = \lambda_c/\lambda_{ab}$, of the superconductor
\cite{Khaykovichetal}.  Here $\lambda_c$ and $\lambda_{ab}$ are the London
penetration depths along the $c$-axis and $ab$-plane, respectively.  In
YBa$_2$Cu$_3$O$_{7-\delta}$ (Y-123), which has a low degree of anisotropy
$\gamma\sim 8$, the vortex lattice melts by a first-order phase transition
into a linelike liquid \cite{Safaretal} up to rather high magnetic fields
$\sim$ 1$-$10 T \cite{Schillingetal}.  In the case of strong anisotropy
$\gamma\sim 500$ and $\sim 150$ as in, respectively, Bi- and Tl-based
high-$T_c$ superconductors a model has been introduced, whereby the
vortices are stacks of two-dimensional ``pancake'' vortices in the CuO$_2$
layers weakly coupled by Josephson and magnetic interactions \cite{Clem}.  At low temperatures and magnetic fields the vortex lattice is composed of
aligned stacks of pancakes (vortex lines).  However, the interactions
between pancakes in adjacent layers are very weak and vortex lines are
easily destroyed by either thermal fluctuations at high temperatures, or
by random pinning.  In Bi$_2$Sr$_2$CaCu$_2$O$_{8+\delta}$ (Bi-2212), which
is the most anisotropic superconductor known, the vortex lattice undergoes
a first-order melting into a pancake liquid at magnetic fields $B\lesssim
500$ G \cite{Khaykovichetal}.

The structure of the vortex liquid phase in the anisotropic
superconductors Y-123 and Bi-2212 has been the focus of
numerous experimental and theoretical studies, e.g. Refs.~
\onlinecite{Safaretal,Shibauchietal,Gaifullinetal1}.  It has been found that at the melting transition in Bi-2212 vortex lines are disintegrated
into a liquid of pancakes \cite{Shibauchietal,Gaifullinetal1}, in
contrast to $\mbox{Y-123}$, where vortex lines are preserved above the melting line
\cite{Safaretal}.  In this Letter we report on the structure of the vortex
state below and near the glass transition in Tl-2212 films, which have an intermediate
anisotropy in between Bi-2212 and
Y-123.  We show that with respect to $c$-axis correlations in the liquid phase
in fields below 0.25~T Tl-2212 films behave as Y-123 with strong disorder
\cite{paul}.

The $c$-axis correlations of pancake vortices are directly related to the
interlayer phase
coherence in the mixed-state \cite{Koshelev}.  One of the most
powerful experimental probes for the interlayer phase coherence in highly
anisotropic layered superconductors is the $c$-axis Josephson plasma
resonance \cite{Gaifullinetal1,Thorsmolleetal}.  The JPR is a Cooper pair
charge oscillation mode perpendicular to the CuO$_2$ layers.  In zero magnetic
field the JPR is a direct probe of the Josephson coupling between the layers
\cite{Gaifullinetal2}.  The JPR frequency is given by:
$\omega_0(T)=c/[\lambda_c(T)\sqrt{\epsilon_{\infty}}]= c/[\gamma\lambda_{ab}(T)
\sqrt{\epsilon_{\infty}}]$.  Here, $c$ is the speed of light, and 
$\epsilon_{\infty}$ is the
high-frequency dielectric constant along the $c$-axis.  In the presence of
a $c$-axis magnetic field $B$, the JPR can be written as \cite{Koshelev}
\begin{eqnarray}
\omega^2_{p}(B,T)=\omega^2_0(T) \langle 
\cos[\varphi_{n,n+1}({\bf r},B)]\rangle.
\end{eqnarray}
Here, $\langle \cos[\varphi_{n,n+1}({\bf r},B)]\rangle$ is the local thermal
and disorder average of the cosine of the gauge-invariant phase difference
between adjacent layers $n$ and $n+1$, and ${\bf r}$ is the in-plane
coordinate.  When the pancake vortices form straight lines perpendicular to
the layers, $\varphi_{n,n+1}({\bf r},B)$ vanishes.  However, when the pancake
vortices are misaligned along the $c$-axis, a nonzero phase difference is
induced, which suppresses the interlayer Josephson coupling, and results in
the reduction of $\langle \cos\varphi_{n,n+1}\rangle$ from unity.  Thus, the
JPR probes the correlations of pancake vortices along the $c$-axis,
providing information on various phases and phase transitions in the
vortex state.

Recent JPR microwave measurements in a Bi-2212 crystal
\cite{Shibauchietal,Gaifullinetal1} show a jump in the interlayer phase
coherence factor, $\langle \cos\varphi_{n,n+1}\rangle$, from $\approx$0.65-0.7
to $\approx$0.3-0.4 at the first-order melting transition.  These data,
together with magnetization measurements, show that the melting of the
vortex lattice is accompanied by disintegration of vortex lines into a
pancake liquid, where $c$-axis correlations are lost.  Quantitative
analysis \cite{Koshelev} of the JPR data confirmed that correlations of
pancakes between neighboring layers in the liquid phase in Bi-2212 far
away from $T_c$ are practically absent.  Only at very low melting fields,
$B_m<B_J=\Phi_0/\lambda^2_J\approx 20$ G, near $T_c$ does the JPR data
\cite{Shibauchietal,Gaifullinetal1} show evidence of a linelike liquid
\cite{Bulaevskiietal1}.  Here $\lambda_J=\gamma s$ is the Josephson length
where $s$ is the interlayer distance.  Vortex lines (stacks of pancake
vortices) are preserved across the melting transition when the intervortex
distance is much larger than the meandering length, $r_w$, defined as the
average in-plane distance between two pancake vortices in adjacent layers
belonging to the same vortex line.

To probe $c$-axis correlations in less anisotropic high-$T_c$ superconductors,
such as thallium and mercury compounds, where the JPR lies in the THz range,
one needs optical techniques.  Demonstrated techniques are either grazing
incidence reflectivity \cite{Tsvetkovetal}, or terahertz
time-domain spectroscopy (THz-TDS) in transmission \cite{Thorsmolleetal}.  Here we report study of the JPR by use of THz-TDS in transmission.  THz-TDS has a
better signal-to-noise ratio, in comparison to grazing incidence reflectivity, allowing measurements close to $T_c$ \cite{Thorsmolleetal}.  Details of the THz-TDS spectrometer are discussed in
Refs.~\onlinecite{Averittetal,Thorsmolle}.  The Tl-2212 film (700 nm) was
grown on a 10 $\times$ 10 mm$^2$ MgO substrate, and exhibited a sharp
transition (0.2 K width) at $T_c = 103.4$ K.  The growth process
is described in Ref.~\onlinecite{Thorsmolleetal}.  The sample was positioned
inside
a cryostat with optical access, between a pair of permanent
magnets with the magnetic field oriented along the $c$-axis.  Measurements
were performed in field cooled mode.  In order to excite the JPR,
$p$-polarized THz radiation, incident at an angle of 45$^\circ$ to the
surface normal, is transmitted through the sample \cite{Thorsmolleetal}.

\begin{figure}[t]
{\centering\resizebox*{0.88\columnwidth}{!}
{\includegraphics{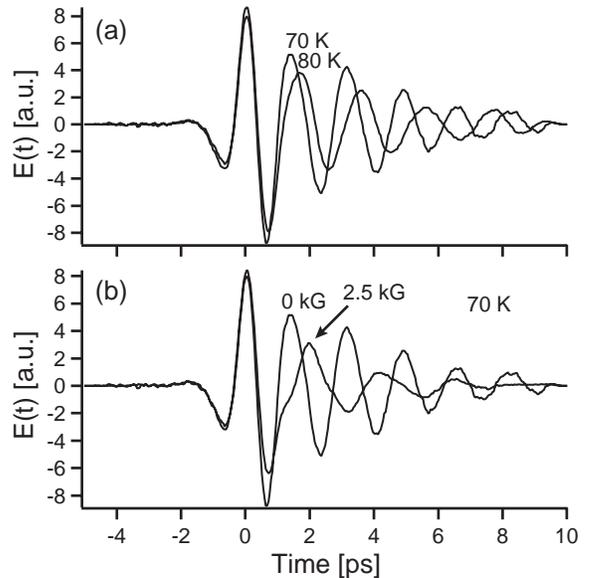}} \par}
\caption{\label{fig: THz pulse JPR w/wo Field} Electric field of THz pulse
in the time-domain transmitted through Tl-2212 thin film at (a) 70~K and 80~K, and at (b) 70 K with and without a 2.5~kG $c$-axis field.\vspace*{-10pt}} \label{JPR THz pulses}
\end{figure}

Fig.~\ref{JPR THz pulses} shows the electric field of the THz pulse
transmitted through the sample.  Fig.~\ref{JPR THz pulses}(a) shows the
oscillations due to the JPR in the time-domain at 70~K and 80~K.  This
clearly illustrates a downward shift in the oscillation frequency as the
temperature ($T$) is increased from 70~K to 80~K.  Fig.~\ref{JPR THz pulses}(b)
shows the JPR in the time-domain at 70~K in zero magnetic field and in a
2.5 kG $c$-axis field.  Here we clearly see the distortion of the THz
pulse by vortices induced by the magnetic field.  It is caused by both diminishing of the
oscillation frequency and broadening of the JPR peak due to the disorder
in pancake positions.

\begin{figure}[t]
{\centering\resizebox*{1\columnwidth}{!}
{\includegraphics{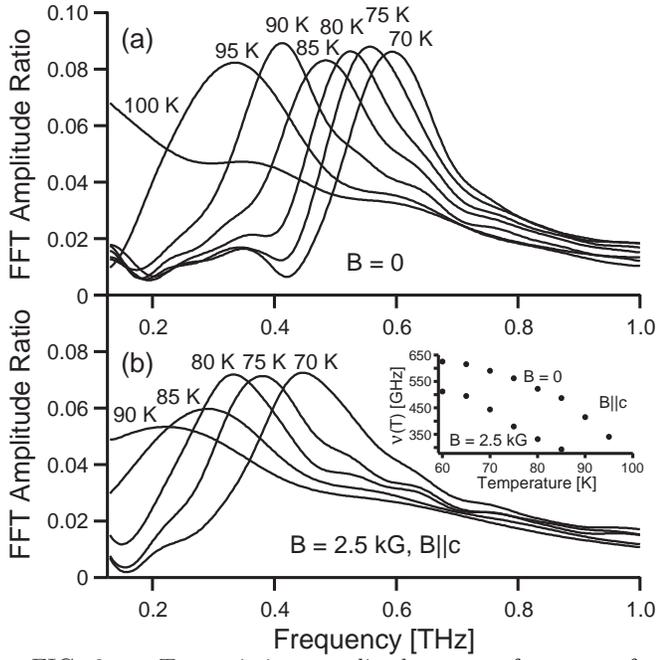}} \par}
\caption{\label{fig: JPR w/wo Field} Transmission amplitude versus
frequency for Tl-2212 thin film.  (a) JPR in zero field from 70 K to 100~K.
(b) JPR in 2.5 kG field applied along the $c$-axis from 70 K to 90 K.  The
inset shows the JPR frequency versus temperature for $B=0$ kG, and $B=2.5$
kG.\vspace*{-10pt}} \label{JPR FFT}
\end{figure}

To obtain the transmission amplitude of the Tl-2212 film, we performed two
sets of averaged scans at each temperature and field.  The first set was
performed on the sample (film plus substrate), and the other set was
performed on a bare reference substrate.  The fast Fourier transform of the
sample was then divided by the fast Fourier transform of the reference.  This
ratio gave the complex transmission coefficient of the Tl-2212 film as a
function of frequency.  Fig.~\ref{JPR FFT} shows the transmission
amplitude as a function of
frequency for the Tl-2212 film (a) in zero field from 70 to 100 K, and
(b) in a 2.5 kG field applied along the $c$-axis.  The inset in Fig.~\ref{JPR FFT}(b) shows the decrease
of the JPR frequency with temperature at $B=0$ and
$B=2.5$ kG, which clearly shows the sensitivity of the JPR to the vortex state.

\begin{figure}[b]
{\centering\resizebox*{1\columnwidth}{!}
{\includegraphics{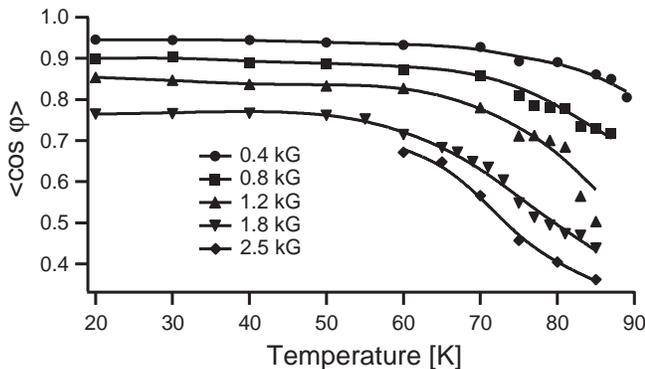}} \par}
\caption{\label{Cosfactor} Interlayer phase coherence factor, $\langle\cos\varphi_{n,n+1}\rangle$, versus temperature in Tl-2212 films at different magnetic fields applied along the $c$-axis.
\vspace*{-10pt}}
\end{figure}

\begin{figure}
{\centering\resizebox*{0.99\columnwidth}{!}
{\includegraphics{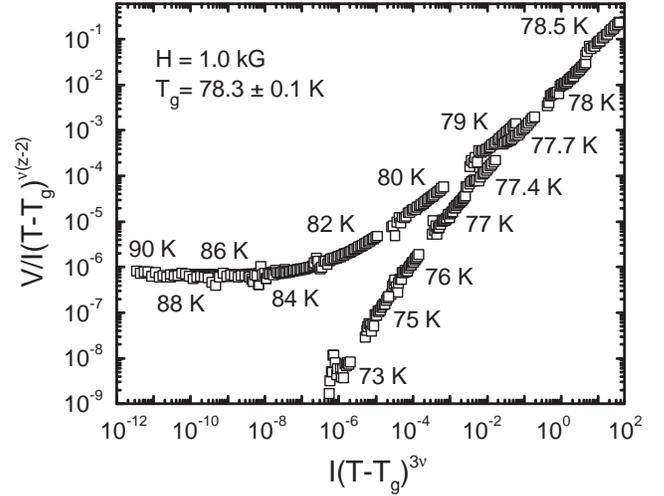}} \par}
\caption{\label{MLT} Scaling of $I$$-$$V$ curves following the Bose-glass
model procedure~\protect\cite{glass}, measured from 73 to 90~K at 1~kG.  We
obtain the exponents $\nu=1.7\pm 0.1$ and $z=4.5\pm 0.2$.  At this field
we find $T_g=78.3\pm 0.1$ K.
\vspace*{-10pt}}
\end{figure}

The interlayer phase coherence factor, $\langle\cos\varphi_{n,n+1}\rangle =
\omega^2_p\left ( B,T \right ) /\omega^2_0\left ( 0, 20\thinspace\mbox{K} \right )$ (determined using data as shown in Fig. \ref{JPR FFT}), as a function of
$T$ at different fields is shown in Fig.~\ref{Cosfactor}.  This factor is practically $T$-independent from
20 to 70~K at $B=0.4$ kG, and from 20 to 50 K at $B=1.8$~kG.  This suggests
that the pancake vortices are well pinned and largely unaffected by
thermal fluctuations at lower $T$ in the vortex solid phase.  $\langle\cos\varphi_{n,n+1}\rangle$ decreases linearly with field at
20~K~$\leq T\leq$~50~K.  Such a field dependence was also observed in the
vortex solid phase in Bi-2212 \cite{Shibauchietal,Gaifullinetal1}.  At
higher $T$ the drop of $\langle\cos\varphi_{n,n+1}\rangle$ with
field becomes stronger, in agreement with theoretical predictions
\cite{Koshelev2}.  We note that Duli\'{c} {\it et al.} \cite{Dulic} did not
observe a change in the field dependence of the JPR frequency,
$\omega_p(B)$, with temperature in grazing incidence reflectivity
measurements up to $70$~K.  The important point is that there is no
indication of a sudden change of the field dependence as a function of
$T$, which would have signaled a disintegration of vortex lines
into a pancake liquid at the melting transition, as observed in
Bi-2212 crystals.

For proper interpretation of our data it is important to determine the transition from the liquid to solid phase and the nature of this
transition (melting or glass).
\begin{figure}[t]
{\centering\resizebox*{1\columnwidth}{!}
{\includegraphics{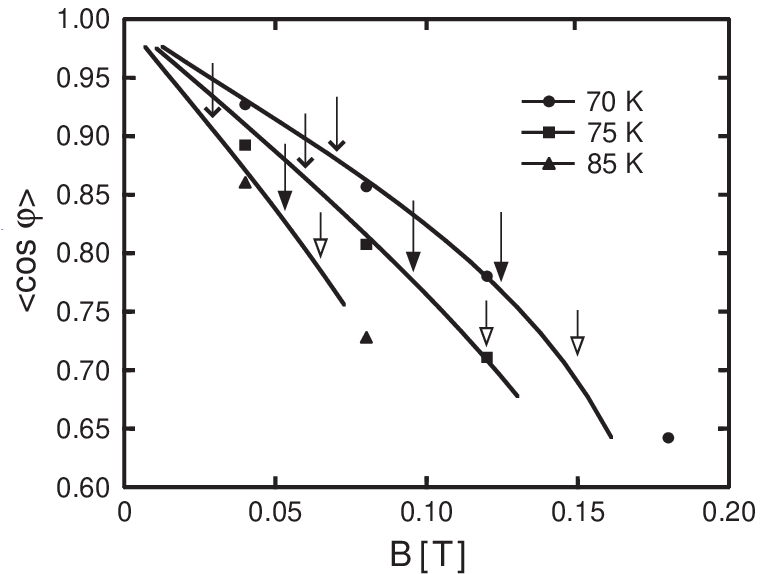}}
\par}
\caption{Comparison between experimental data (symbols) for the interlayer
phase coherence factor, $\langle\cos\varphi_{n,n+1}\rangle$, in Tl-2212 films at
70, 75 and 85~K, and numerical calculations (solid lines) based on
Ref.~\protect\cite{Koshelev2}.  Calculations were done using $\lambda_c$
extracted directly from the JPR frequency with $\epsilon_{\infty}=9.1$ and
assuming the anisotropy factor $\gamma=150$.  Arrows mark glass
transitions determined from transport measurements (open arrows) and estimated
melting transition (closed arrows).  Simple arrows indicate irreversibility line in single crystals \protect\cite{HardyPRB97}.
\vspace*{-10pt}}\label{fig: JPR vs Fit}
\end{figure}
To address this issue we performed $I$$-$$V$ measurements in fields from
0.5 to 1.5 kG as a function of $T$.  Fig.~\ref{MLT} shows the
scaled $I$$-$$V$ curves in 1.0 kG.  We determined the glass
transition temperature $T_g$ by scaling the nonlinear $I$$-$$V$ curves via
the Bose-glass model \cite{max}.  When scaled, the $I$$-$$V$ data fall
onto different curves for temperatures above and below $T_g$.  The positions
of $T_g$ determined in this manner at 70, 75, and 85~K are
depicted in Fig.~\ref{fig: JPR vs Fit}.  Hence, in our films the melting
transition is replaced by the glass transition due to disorder inherent to
the films.  Paulius {\it et al} \cite{paul} observed such replacement in
Y-123 crystals under the effect of controlled disorder introduced by
proton irradiation.  For Tl-2212 single crystals, the irreversibility line
was observed from magnetization measurements by Hardy {\sl et al.}
\cite{HardyPRB97} at fields of 0.7, 0.6 and 0.3 kG at
$T=70$, 75 and 85~K, respectively (see Fig. \ref{fig: JPR vs Fit}).  In films the melting
or glass transition field is strongly enhanced by pinning as compared to single crystals.

Next we address whether we have linelike vortices in the solid and
liquid phases and how these phases differ with respect to $c$-axis
correlations.  For this we discuss the field dependence of the coherence factor
$\langle\cos\varphi_{n,n+1}\rangle$ below and through this transition.

Suppression of the Josephson coupling due to pinning-induced meandering of
vortex lines at low $T$ has been considered in
Ref.~\protect\cite{Koshelev3}.  Let us discuss the isolated vortex regime
at low fields $B\ll \Phi_0/4\pi\lambda_{ab}^2, B_J$, i.e. well below 0.05
T.  Here the coherence factor drops linearly with $B$,
\begin{equation}
\langle\cos\varphi_{n,n+1}\rangle = 1-\frac{B}{B_w}, \ \
B_w\approx\frac{\Phi_0}{\pi r_w^2\ln(\gamma s/r_w)},
\end{equation}
where the field $B_w$ is a measure of the meandering length $r_w$.  $r_w$, 
in turn, is determined by a balance between the pinning
energy $U_p$ and the Josephson energy, $E_J=\Phi_0^2/16\pi^3s\lambda_c^2$,
$r_w^2\sim U_p/E_J$.  Our data, Fig.~3, give
$B_w\approx 8$~kG corresponding to $r_w\approx 25$ nm.  This value for
$B_w$ has to be compared with the lower value $\approx 1.2$ kG for the
more anisotropic Bi-2212 \cite{Gaifullinetal1}, the similar value for a
different Tl-2212 film, $\approx 14$ kG \cite{Dulic}, and with the much higher
value $\approx 30$ T for the less anisotropic underdoped
Y-123 \cite{Dulic}.  This shows that $B_w$ is mainly
determined by the anisotropy of the material.  In the isolated vortex
regime we found an intervortex distance $a=\sqrt{\Phi_0/B}>$ 200 nm.  Hence, disorder does not
destroy vortex lines.

At higher temperatures, meandering of vortex lines is determined by thermal fluctuations.  We calculate the $T$- and $B$-dependence of
the interlayer phase coherence factor, $\langle\cos\varphi_{n,n+1}\rangle$,
in the vortex solid phase neglecting pinning but accounting for Gaussian
thermal fluctuations of pancakes \cite{Koshelev2}, accounting for both
Josephson and magnetic coupling of the layers.  Calculations are based
on the Lawrence-Doniach model in the London limit, which is completely
defined by the London penetration depths $\lambda_{c}$ and $\lambda_{ab}$,
and the interlayer spacing, $s=15$~\AA.  We extract
$\lambda_{c}$ ($\approx 0.003$ cm at 70 K) directly from the JPR frequency
using $\epsilon_{\infty}=9.1$ \cite{DulicPRB99}.  We find that 
the best agreement
between calculation and experiment is achieved using $\gamma=150$, see 
Fig.~\ref{fig: JPR vs Fit}.  This
value of $\gamma$ is consistent with the resistivity anisotropy of Tl-2212 
single
crystals near $T_c$ \cite{HardyPRB97}. 

Extracted parameters of our film allow us to calculate the \emph{expected}
location of the melting transition at different fields for a corresponding clean
material.  In superconductors with dominating Josephson coupling, $\gamma
<\lambda /s$, the melting temperature is given by
$T_{m}\approx0.13\epsilon_{0}a/\gamma $ with
$\epsilon_{0}=\Phi_{0}^{2}/\left ( 4\pi\lambda_{ab} \right )^{2}$ 
\cite{Koshelev-Nordborg:PRB99}.  In our case the anisotropy
factor $\gamma\approx 150$ is comparable with the ratio $\lambda_{ab}/s$
and the magnetic coupling between pancake vortices in neighboring layers
may influence the melting transition.  We calculated the shift of the
melting transition caused by magnetic coupling using perturbation theory 
and obtain \vspace*{-5pt}
\begin{equation}
T_{m}\left ( B \right ) \approx 0.13(\varepsilon_{0}a/\gamma)(1+
0.23a^{2}/\lambda_{ab}^{2}). \vspace*{-8pt}\label{TmCorr}
\end{equation}
The position of both the \emph{estimated} $T_m$ (closed arrows) and \emph{experimentally determined} $T_g$ (open arrows) is indicated in Fig.~\ref{fig: JPR vs Fit}.  We find that the glass transition fields exceed the calculated melting fields in a clean material by $\sim$15$-$20\%, while the irreversibility fields are even smaller in a single crystal.  It is generally true that irreversibility is resolution limited and lies lower than $T_m$.  It has been established that
\emph{point} disorder drives the glass transition \emph{below} the melting line \cite{paul}, while \emph{correlated} disorder drives it
\emph{above} the melting line \cite{paulCorr:PRL00}.  Therefore, our
results suggest that pinning in our films is dominated by correlated
disorder, presumably dislocations \cite{Dam:Nat99}.  This is in contrast to single crystals where point disorder is likely to dominate.  In our films, at glass temperatures
in the isolated vortex regime in the solid phase, $B_w$ decreases about
two times in comparison with low temperatures and $r_w$ increases to 35~nm
at 80~K.  The field dependence is completely smooth near the glass
transition.  These results strongly imply that vortex lines are preserved
across the glass transition for $T\geq 70$~K.
%

In conclusion, we have measured the JPR in Tl-2212 thin films using THz-TDS
in transmission.  We used the JPR frequency measurements as a probe to study
the interlayer phase
coherence as a function of temperature and magnetic fields above and below
the melting line.  Our results show that in Tl-2212 films the glass transition
takes place probably due to disorder inherent to the films.  Vortex
lines exist in the liquid state in magnetic fields below 2.5 kG.
We conclude that Tl-2212 films, at least in the
low magnetic field portion of the phase diagram, behave as Y-123, and not
as Bi-2212 single crystals despite the closer anisotropy ratios of Bi-2212 and Tl-2212.

We thank Superconductivity Technologies for providing the Tl-2212 films.  This research was supported
by the UC Campus-Laboratory Collaborations and by
the Los Alamos LDRD Program by the U.S. DOE.  Work in Argonne was supported by the U.S. DOE,
Office of Science, under contract No. W-31-109-ENG-38.

%
%


%
%

\end{document}